\documentclass[reprint,aps,groupedaddress,showpacs,%
amsmath,preprintnumbers,prl,nofootinbib]{revtex4-1}

\usepackage{graphicx}
\usepackage{amsmath}
\usepackage{amsfonts}
\usepackage{amssymb}
\usepackage{amsbsy}

\newcommand*{\eff}{\text{eff}}

\usepackage{colordvi}

\newcommand{\Ham}{\mathcal{H}}
\newcommand{\Hil}{\mathcal{H}}
\newcommand{\ket}[1]{{|#1\rangle}}
\newcommand{\N}{\mathcal{N}}
\newcommand{\re}{\mathbb{R}}
\newcommand{\Pl}{{\rm Pl}}
\newcommand{\lPl}{\ell_{\Pl}}

\def\p {\partial}

\newcommand{\rd}{{\rm d}}

\begin{document}

\title{Time and a physical Hamiltonian for quantum gravity}
\author{Viqar \surname{Husain}${}^1$}
\email{vhusain@unb.ca}
\author{Tomasz \surname{Paw{\l}owski}${}^{1,2}$}
\email{${}^2$ present address; tpawlow@fuw.edu.pl}

\affiliation{
  ${}^1$ Department of Mathematics and Statistics, University of New Brunswick, Fredericton, NB, Canada E3B 5A3
  \\
  ${}^2$ Katedra Metod Matematycznych Fizyki, Uniwersytet Warszawski, ul. Ho\.{z}a 74, 00-681 Warszawa, Poland  
} 


\begin{abstract}

We present a  non-perturbative quantization of general relativity coupled to dust and other matter fields. The dust provides a natural time variable, leading to a physical  Hamiltonian with spatial diffeomorphism symmetry. The surprising feature  is that the Hamiltonian is not a square root. This property,  together with the kinematical structure of loop quantum gravity, provides a complete theory of quantum gravity, and puts in technical reach applications to  cosmology,  quantum gravitational collapse and Hawking radiation.

\end{abstract}

\maketitle

The problem of finding a quantum theory of gravity has been a central one in theoretical physics
for several decades. There are two main approaches  to the problem with rather different starting points. The particle physics approach is one  in which Lorentz invariance and its associated background structure is a central axiom, and remains one in perturbative string theory.   The non-perturbative canonical approach originates in the view that general relativity and related theories are geometric in nature and should be treated without a break up into a fixed background and a metric perturbation on it.

The non-perurbative approach began with the canonical formulation of general relativity due to Arnowitt, Deser and Misner (ADM) \cite{adm}. This  was used by DeWitt \cite{dewitt} to formulate  quantization in the metric representation, and led to the Wheeler-DeWitt equation and as a  being a central one of quantum gravity. Subsequent work on quantum geometrodynamics (QGD) in the 1970s by Misner \cite{Misner}, Kuchar \cite{Kuchar} and others, and in the early 1980s by Hawking and Hartle \cite{HH} focused on finding the quantum theories of symmetry reductions of general relativity.

Since the mid-1980s the non-perturbative approach has evolved into the loop quantum gravity
program, where the ADM canonical variables are replaced by the triad and connection introduced by Ashtekar \cite{aa}. This has since developed into a mathematically well-defined formulation in which the constraints of general relativity are realized as  operators on a kinematical Hilbert space   \cite{al-status} . With regard to dynamics,  there are several developments in mini-superspace models that have given new results specifically with  applications to cosmology and singularity avoidance \cite{mb-lqc, aa-lqc}. There has also been progress on midi-superspace models and gravitational collapse using LQG ideas  \cite{gowdy,coll-vh,coll-zip}. However a full theory of quantum gravity without symmetry reductions remains elusive.

Among the obstacles in formulating a non-perturbative theory of quantum gravity is the 
problem of time and its associated conserved inner product. The problem arises because of the time reparametrization invariance of the theory. This invariance is manifested in the canonical theory by a Hamiltonian constraint rather than a  non-vanishing physical Hamiltonian.

Although any theory can be made locally time gauge invariant, such ``parametrization'' is an artificial construct  that gives a Hamiltonian constraint linear in  one of the canonical momenta. This linearity in turn provides a natural time gauge fixing which allows the recovery of the original un-parametrized theory.  The problem is that general relativity coupled to usual matter is not  a parametrized theory  in this sense, since the Hamiltonian constraint is quadratic in all momenta.

A ``solution'' to the problem of time could be to make a suitable time gauge fixing in the classical theory and obtain the corresponding true Hamiltonian. This leads to two generic problems:  (i) different time gauges give different Hamiltonians with no way known to connect them (unlike in Minkowski spacetime and other background dependent theories) where  the Lorentz  transformations
are represented on the physical Hilbert space and connect different frames, and (ii) physical  Hamiltonians have a form that is a square root. This is because the Hamiltonian constraint is quadratic or possibly worse (depending on matter potential energy terms) in all canonical variables, so that solving for the Hamiltonian after gauge fixing is at best a quadratic equation.

A solution to the problem of time is suggested in work of  Brown and Kuchar \cite{bk} who introduced a pressureless dust coupled to general relativity. Their action leads to a remarkable canonical theory in which the Hamiltonian constraint is linear in the dust momentum. This suggest a natural ``theory provided'' time variable and reduced Hamiltonian. Furthermore the dust is such that it provides a  {\it time and space reference system}. However, if the dust frame is used to gauge fix the classical theory, the reduced hamiltonian is still a square root. This approach is studied in the LQG  context in \cite{Giesel-Thiemann} where a fully reduced quantization is suggested, but this obstacle remains unresolved beyond a formal level. (Other  works with scalar field time have the same problem \cite{RS-93,dgkl-qg}). 

In this paper we provide a solution to both the problem of time and that of the square root by introducing a  modification of the Brown-Kuchar dust and  combining it with the kinematical results of LQG.  The approach uses a ``hybrid'' quantization in the sense that only a time gauge is fixed, but the spatial coordinate gauges are not;  therefore the spatial diffeomorphism constraint remains in the theory.  Combined with the kinematical results of LQG, this approach leads to a complete and rigorous formulation of a quantum gravity theory.  The Hamiltonian of the theory {\it is} what would be the Hamiltonian constraint without the dust, and the kinematical Hilbert space of LQG becomes (up to spatial diffeomorphisms) the physical Hilbert space. This space  carries a unitary representation of spatial diffeomorphisms \cite{almmt}, so the resulting picture is like that of the Poincare group carried  on the Hilbert space of  quantum field theory. 

The theory we consider is given by the action
\begin{equation}
  \begin{split}
    S &= \frac{1}{4G} \int \rd^4x \sqrt{-g} R \ +\ S_{\rm SM} \\
      &\hphantom{=} - \frac{1}{2} \int  \rd^4x\  \sqrt{-g} M ( g^{ab}\partial_a T \partial_b T + 1) ,  
  \end{split}
\end{equation}
where $S_{\rm SM}$ is the action for any standard model matter.
In addition to the metric $g_{ab}$ and  matter fields in $S_{\rm SM}$, this action contains
the dust field $T$, with $M$ enforcing its gradient to be timelike. The last term
resembles the Brown-Kuchar dust  action \cite{bk}, but is different in that the dust is 
irrotational. {(The action with one scalar was also considered as a possible simplification in \cite{bk,kt}, but its full possibilities at the quantum level were not explored).} 
It also resembles the so-called ether models studied in \cite{ether}, where the dust four velocity
  is not written as the gradient of a scalar, and contains other dynamical terms. 
With $U_a = \p_aT$ the dust stress-energy tensor is
\begin{equation}
  T^{ab} = M U^a U^b + (M/2) g^{ab}\left( g_{cd} U^cU^d +1 \right);
\end{equation}
this is the usual form of dust field with rest mass $M$.

Since the Hamiltonian theory of the gravity coupled to matter is well-known
in both the ADM and Ashtekar-Barbero variables, we only need to obtain the canonical
formulation of the dust. Substituting  the ADM form of the metric
\begin{equation}
  \rd s^2  = -N^2 \rd t^2 + q_{ab}(N^a\rd t + \rd x^a)(N^b\rd t +\rd x^b)
\end{equation}
in the dust lagrangian gives
\begin{equation}
  L_D = \frac{M\sqrt{q}}{2N} \left[ ( \dot{T} + N^a \p_aT )^2 - N^2(q^{ab}\p_aT \p_bT +1)\right]
\end{equation}
The dust momentum is
\begin{equation}
  p_T = \frac{\p L_D}{\p \dot{T}}= \sqrt{q} \frac{M}{N} (\dot{T} + N^a\p_aT ),
\end{equation}
which gives the canonical action
\begin{equation}
  S_D = \int \rd t \rd^3x\left[ p_T\dot{T} -N {\cal H}_D - N^a C_a^D  \right],
\end{equation}
where
\begin{subequations}\begin{align}
  {\cal H}_D &=  \frac{1}{2}\left[ \frac{p_T^2}{M\sqrt{q}}  + \frac{M\sqrt{q}}{p_T^2}\left(p_T^2 +q^{ab}C_a^DC_b^D\right)  \right]\\
  C^D_a &=   -p_T \p_aT.
\end{align}\end{subequations}
Now the equation of motion for $M$ gives
\begin{equation}
  M =  [q]^{-1/2}\, p_T^2\, [p_T^2 + q^{ab}C_a^DC_b^D]^{-1/2}.
\end{equation}
Substituting this back into the dust canonical action yields 
\begin{equation}
  S_D = \int \rd^3x\ \rd t \left[ p_T \dot{T} -N \sqrt{p_T^2 + q^{ab}C_a^DC_b^D} + N^a C_a^D\right].
\end{equation}

At this stage it is evident from this action that the dust contribution to the full Hamiltonian constraint is the square root in the last equation, with the expected addition to the spatial diffeomorphism constraint. The difference from the {main} Brown-Kuchar  result is that $C_a^D$ has a simpler form, being composed of only one scalar $T$ and its conjugate momentum. This is crucial for what follows.

Let us now impose the canonical time gauge fixing condition  $T=t$. This is of course  the obvious choice for the parametrized particle and scalar field, and is also natural here; {(this gauge is also discussed in \cite{kt})}. This condition is second class with the Hamiltonian constraint
\begin{equation}\label{eq:hc1}
  {\cal H} =  \sqrt{p_T^2 + q^{ab}C_a^DC_b^D} + {\cal H}_G + {\cal H}_{\rm SM}=0,
\end{equation}
where the last two terms are the standard gravitational and matter contributions
to this constraint. Substituting the gauge condition into this constraint gives the physical 
Hamiltonian density 
\begin{equation}
  \tilde{H} := -p_T =  {\cal H}_G + {\cal H}_{\rm SM}, \label{trueH}
\end{equation}
which is just the sum of the gravitational and non-dust matter energy densities. 
{(We note the same Hamiltonian density may be obtained by using the strongly commuting
Brown-Kuchar Hamiltonian constraint \cite{bk}, with one scalar non-zero and the same gauge choice.)}

The spatial diffeomorphism constraint  takes the form
\begin{equation}
  {\cal C}_a \equiv C^G_a + C^{\rm SM}_a = 0
\end{equation}
in this time gauge, since $p_T\p_aT= p_T\p_a t=0$. ($C^G_a$ and $C^{\rm SM}_a$ are
the usual gravitational and matter parts of this constraint.) Lastly we note that
the physical Hamiltonian  is (spatial) diffeomorphism invariant: $ \{ {\cal C}(N), \int \rd^3x\ \tilde{H}\}  = 0$, 
where ${\cal C}(N) = \int_\Sigma \rd^3x\ N^a {\cal C}_a$, and the diffeomorphism constraint algebra is, as expected,  
 $ \{ {\cal C}(N), {\cal C}(M) \} = {\cal C}([N,M])$. 

The spacetime metric is obtained by requiring that the condition $T=t$ is dynamically 
propagated: $\dot{T} = \dot{t} = 1 = \{ T, \int_\Sigma \rd^3x (N{\cal H} + N^a {\cal C}_a)  \} |_{T=t}$.
This gives lapse $N=1$ but  leaves the shift $N^a$ unconstrained.  The gauge fixed Hamiltonian action
in the ADM canonical variables $(q_{ab}, \tilde{\pi}^{ab})$ is therefore 
\begin{equation}
S^{GF} = \int \rd^3x \rd t \left(\tilde{\pi}^{ab} \dot{q}_{ab} - \tilde{H} - N^a \mathcal{C}_a\right), 
\end{equation}
and {\it time reparametrization is no longer a gauge symmetry}.   
The theory  formally resembles a Yang-Mills gauge theory, but with the  Gauss law  replaced by the spatial diffeomorphism constraint. Furthermore the physical Hamiltonian $\tilde{H}$ is not a square root,  a fact that removes a fundamental hurdle for quantization. The term ${\cal H}_G$ in  $\tilde{H}$ (\ref{trueH})  may be written in either the ADM  or
in the triad-connection variables. We will consider the latter  where there exists a well
developed kinematical quantization \cite{almmt} {leading to the space $\Hil_{\rm diff}$ of diffeomorphism invariant states.} 
With our framework {\it this carries over unaltered, but becomes the physical quantization.}

In the triad-connection variables the canonical phase variables are the pair $(A_a^i, E^a_i)$ where
$A_a^i=\Gamma^i_a(E) + \gamma K_a^i$ is an su(2) connection and $E^{ai}$ is a vector density of weight one. They satisfy the Poisson bracket
\begin{equation}
  \{  A_a^i, E^b_j\} = \delta^i_j \delta^a_b \delta(x,y).
\end{equation}
The gravitational part of the Hamiltonian (\ref{trueH}) is 
\begin{equation}\label{eq:Ham-class}
  {\cal H}_G = \frac{\gamma^2}{2 \sqrt{{\rm det} E}}E^a_iE^b_j\left(\epsilon^{ij}_{\ \ k} F_{ab}^k + 2(1-\gamma^2)
  K_{[a}^iK_{b]}^j \right),
\end{equation} 
withe the variables subject to the Gauss and spatial diffeomorphism constraints
\begin{subequations}\label{eq:constr-class}
\begin{align}
  {\cal G}_i &= \p_a E^a_i + \epsilon^k_{\ ij} A_a^j E_k^a, \label{eq:sonstr-class-gauss} \\
  C_a^G &=  E^b_i F_{ab}^i - A_a^i{\cal G}_i. \label{eq:constr-class-diff}
\end{align}
\end{subequations}

The phase space variables used for quantization are the holonomy
$U_\gamma(A) \equiv P\exp \int_\gamma A_a^i \tau^i \rd x^a$ and the flux
$K^i = \int_S E^{ai}\rd\sigma_a$, where the loop
$\gamma$ and surface $S$ are  embedded in a spatial slice, and $\tau^i$ is a generator of the group. These variables  satisfy the Poisson bracket 
\begin{equation}
\{ U_\gamma, K^i \} = \int_\gamma \rd s \int_S \rd\sigma_a \dot{\gamma}^a(s) \delta^3(\gamma(s),S(\sigma)) \tau^i U_\gamma
\end{equation}   
This algebra is quantized on the  Hilbert space {$\Hil_{\rm kin}$} spanned by the spin-networks states. The basis is labelled by three sets of quantum numbers: a graph embedded in a 3-manifold, 
by an assignment of spin labels on its  edges, and by  intertwiners on its vertices (which 
sew together the spins entering a vertex) \cite{Thiemann-book}.  

The main results of this kinematical quantization (also known as {\it polymer quantization}) are:  
(i) the representation is unique \cite{LOST} and background independent, 
(ii) the {solutions to} the Gauss law \eqref{eq:sonstr-class-gauss} {form the subspace of $\Hil_{\rm kin}$ spanned by
spin network with trivial intertwiners},
(iii) implementation of the quantum (spatial) diffeomorphism constraint is well understood 
\cite{almmt}, 
(iv) geometric area and volume operators {are well defined on the resulting space $\Hil_{\rm diff}$ 
and} have discrete spectra, and lastly 
(v) the Hamiltonian constraint can be {also} written as a  well-defined operator on {$\Hil_{\rm diff}$}. 

The problem not solved in LQG is the determination of the physical Hilbert space and observables, a step  which can only be completed  by solving the quantum Hamiltonian constraint. {\it It is this final problem that is bypassed in our approach:} a complete formulation of a quantum theory of gravity
results because (v) gives the physical Hamiltonian $\hat{\Ham}_G$ of the present theory. This step also resolves  the square root obstacle in previous attempts to construct a deparametrized quantum gravity, which made the resulting formalism difficult to apply, even for homogeneous models with arbitrary matter fields.

There are two approaches for formulating the physical Hamiltonian $\hat{\Ham}_G$.
In its original form, it acts   by adding particular edges to the graph of a spin-network basis state. 
 In an alternative formulation, known as \emph{algebraic} LQG 
\cite{aqg}, $\hat{\Ham}_G$ leaves the graph unchanged, but modifies
the other quantum numbers. This makes it technically easier for 
physical applications. {It is this latter approach that we apply in our formulation.}

{To define the action of the Hamiltonian one first chooses a fixed graph $\alpha$. The 
choice includes (but is not restricted to) the triangular or cubic lattice. The operator itself has the form of the sum over the graph vertices 
$\hat{H}^G=\sum_{v\in V(\alpha)}\hat{\mathcal{H}}^G_v$, where $\hat{\mathcal{H}}^G_v$ is 
composed of $(i)$ the volume operator $\hat{V}(v)$, $(ii)$ the combination $\hat{h}_e[\hat{h}_e^{-1},\hat{V}]$ with holonomies 
$\hat{h}_e$ along the adjacent edge, and $(iii)$ the holonomies $\hat{h}_{\square}(v)$ along
the minimal closed loops in $\alpha$. The operators $(i)$ and $(ii)$ are diagonal and their properties 
are well understood (see \cite{b-v} for $\hat{V}$) whereas $(iii)$ changes the spin labels on the edges composing the loop.}

{We note that  because we have a physical Hamiltonian which comes with a  uniquely fixed lapse,
there is no anomaly in the quantum theory, unlike the formulation in Ref. \cite{aqg}. This is because now the commutator of the Hamiltonian with itself vanishes identically.}  

 The coupling of matter and its polymer quantization are also well understood. The physical Hilbert space is extended  by adding  matter quantum numbers on the vertices and edges  of graphs, depending on the type of matter. The form and the action of the quantum counterparts of the matter Hamiltonian (and  matter part of the diffeomorphism constraint) is also explicitly known \cite{Thiemann-book}. 

The low energy consequences of the theory can be probed by  semiclassical states and observables. Here again the needed elements are at our disposal:  the  kinematical coherent states and diffeomorphism-invariant observables  are available \cite{coh,Thiemann-book}, and immediately become physical in our approach. Thus there is no need to utilize the partial observable formalism \cite{obs-r,*obs-d}.

 {The development presented here brings together three aspects: $(i)$ natural time 
gauge fixing, $(ii)$ simplification of the (physical) Hamiltonian by elimination 
of the square root, and $(iii)$ the diffeomorphism-invariant framework of LQG. It is the unique
combination of these elements which allows a completion of the gravity quantization program.} 
{Details of this  approach are to appear in a forthcoming paper \cite{lqg-det}.}

{This development} has applications to a spectrum  of problems that require a quantum theory of gravity. We discuss three possibilities: cosmology, gravitational collapse and Hawking radiation, and quantum field theory in curved spacetime. Details of the first  have already appeared \cite{HP-cosm}, and the second is in preparation. 

\noindent\underbar{Cosmology:} 
The simplest application is to homogeneous and isotropic models.  These have been much studied 
in standard LQC, where a scalar field is used as a clock. This  has significant limitations because it cannot be extended to arbitrary scalar field potential or to  other types of matter. These issues are overcome in our approach, which in  addition provides significant technical simplicity for both polymer and Schrodinger quantization.

For the isotropic flat model, described by the metric $\rd s^2 = -N^2(t)\rd t^2 + a^2(t)(\rd {\rm x}^2+\rd {\rm y}^2+\rd {\rm z}^2)$, the Hilbert space  reduces to $L^2(\bar{\re},\rd\mu_B)$, where $\bar{\re}$ is the Bohr compactification  of the real line. Its basis is provided by eigenstates $|v\rangle$ where  
$v\propto a^3$. The diffeomorphism constraint vanishes identically and the 
Hamiltonian becomes
\begin{equation} 
  \hat{\Ham}_{G} = -\frac{3\pi G}{8\alpha} \sqrt{|\hat{v}|} (\hat{\N}-\hat{\N}^{-1})^2 \sqrt{|\hat{v}|}   
   + \frac{3\rho_c}{16\alpha}\Lambda|v|,
\end{equation}
where $\hat{\N}\ket{v}=\ket{v+1}$ and $\alpha\approx 1.35\lPl^3$, $\rho_c\approx 0.82\rho_{\Pl}$ are 
constants. This Hamiltonian is essentially self-adjoint and its spectrum can be found
analytically.   Explicit  unitary evolution  provides a  controllable way of studying the 
dynamics in \emph{all} epochs; in particular the big bang singularity is replaced by
a bounce.  The analysis is readily  extended for coupling to a scalar field or other  matter, since polymer realizations of the matter Hamiltonians  are available.  Details of these results  appear in \cite{HP-cosm}. 
 
For  inhomogeneous cosmology, both for metric fluctuations about FRW or non-perturbative
models such as Gowdy or cylindrical waves,  our approach benefits  from the existence of a (manageable) physical  hamiltonian and a simple and  anomaly free algebra of constraints. For  cosmology  the gauge invariant  linear perturbation
theory  \cite{l-pert} is  directly applicable and technically manageable; it has the potential  to provide an  observational signature.

For nonperturbative models, the approach  again provides a convenient starting point  for the implementation of midi-superspace/hybrid quantization schemes \cite{mb-lqc,gowdy}. It has the potential   to address directly the existing difficulties of these treatments, such as the problem of renormalizability  in the inhomogeneous degrees of freedom.

\noindent\underbar{Gravitational collapse and Hawking radiation:}   The classical problem is  well understood in spherical symmetry with a minimally coupled scalar field \cite{chopt},  where critical behaviour at the onset of black hole formation and a finely tuned violation of the cosmic censorship conjecture (CCC) are observed.  This model is much richer than CHGS, which has only pure inflow and  a black hole always forms, so there is no critical behaviour. Quantization of the model is important to study for at least three reasons: (i)  the expectation of singularity avoidance in quantum gravity must modify the CCC violating critical solution, (ii) the presence of both  inflow and outflow can qualitatively affect  Hawking radiation, and (iii) a unitary quantum theory would immediately solve the information loss problem.  
 
Our approach provides two concrete models, one with dust only, and the other with dust and scalar field. Both are midi-superspace models which can investigated  in either Ashtekar-Barbero or ADM variables using polymer quantization.  The spatial metric is parametrized as $\rd s^2= \Lambda^2(r,t) \rd r^2 + R(r,t)^2 \rd\Omega^2$,  and there is a radial diffeomorphism constraint and a physical Hamiltonian. The first model  has one local degree of  freedom and the second  has two.  At the quantum level  the models are  technically straightforward to write down, unlike the unmanageable  non-local Hamiltonian without dust  first discussed by Unruh \cite{U-nlH}.  

The quantum theory of the dust + scalar field in spherical symmetry is presently under study by the authors. There is already strong indication that a self-adjoint physical Hamiltonian is available, a feature intimately linked  with resolving the information loss paradox. 
 
\noindent\underbar{Quantum field theory on curved spacetime}  Another expectation of 
a quantum theory of gravity is that it provides a low energy ``emergent'' quantum field theory on
a curved background. In our framework  this is accomplished by first choosing a  {\it dynamical}
semiclassical state of geometry $|\psi(\tilde{g}) \rangle_G$ peaked on a classical trajectory $\tilde{g}$
that provides  the desired background. (As we have noted, such states are available since  we have solved the problem of time, and the Hilbert space is the physical one.)  The state is then used to obtain an effective background dependent  Hamiltonian
\begin{equation}
 \hat{ \Ham}^{\eff} = \  _G\langle \psi(\tilde{g})|\hat{\Ham}_{G} |\psi(\tilde{g}) \rangle_G\  \hat{I} + \hat{\Ham}_m(\tilde{g}).
\end{equation}
 This Hamiltonian acts only on the matter Hilbert
space,  so this procedure yields  a  QFT on curved spacetime. Thus a given 
semiclassical state for the  gravity  sector provides an unambiguous matter vacuum obtained
by finding the ground state of this effective Hamiltonian.   

In contrast,  the standard  semiclassical approximation, $G_{ab} = 8\pi \langle T_{ab}\rangle$,  
requires first selecting a  matter state (in Fock quantization)  in which the stress-energy tensor expectation value is computed,  and then solving this equation for the metric. This is the arena 
for the usual Hawking radiation calculation. 
   
 In summary, we have presented  a complete and computationally accessible theory of
quantum gravity, and outlined  a wide range of applications, from cosmological models to gravitational collapse. 
  
\begin{acknowledgments}
  This work was supported by the Natural Science and Engineering Research Council of Canada.
\end{acknowledgments}

\vspace{-3mm}

\bibliographystyle{apsrev4-1}
\bibliography{qg-letter.short}

\end{document}